# Towards metropolitan free-space quantum networks

Andrej Kržič[1,2,*], Sakshi Sharma[1,2], Christopher Spiess[1,2], Uday Chandrashekara[1,2], Sebastian Töpfer[1], Gregor Sauer[1,2], Luis Javier González-Martín del Campo[1,2], Teresa Kopf[1], Stefan Petscharnig[3], Thomas Grafenauer[3], Roland Lieger[3], Bernhard Ömer[3], Christoph Pacher[3,4], René Berlich[1], Thomas Peschel[1], Christoph Damm[1], Stefan Risse[1], Matthias Goy[1], Daniel Rieländer[1], Andreas Tünnermann[1,5,6], Fabian Steinlechner[1,6,†]

[1] *Fraunhofer Institute for Applied Optics and Precision Engineering, Albert-Einstein-Str. 7, 07745 Jena, Germany*
[2] *Friedrich Schiller University Jena, Faculty of Physics and Astronomy, Max-Wien-Platz 1, 07743 Jena, Germany*
[3] *AIT Austrian Institute of Technology, Giefinggasse 4, 1210 Vienna, Austria*
[4] *fragmentiX Storage Solutions GmbH, IST Park, Plöcking 1, 3400 Klosterneuburg, Austria*
[5] *Institute of Applied Physics, Friedrich Schiller University Jena, Albert-Einstein-Str. 15, 07745 Jena, Germany*
[6] *Abbe Center of Photonics, Friedrich Schiller University Jena, Albert-Einstein-Str. 6, 07745 Jena, Germany*

Quantum communication has seen rapid progress towards practical large-scale networks, with quantum key distribution (QKD) spearheading this development. While fibre-based systems have been shown to be well suited for metropolitan scales, suitable fibre infrastructure may not always be in place. Here, we make the case for an entanglement-based free-space quantum network as a practical and efficient alternative for metropolitan applications. We developed a deployable free-space QKD system and demonstrated its use in realistic scenarios. For a representative 1.7-km free-space link, we showcase its ad hoc deployability and achieve secure key rates of up to 5.7 kbps, with 2.5 kbps in direct noon sunlight. By extrapolating experimental data, we show that kbps key rates are achievable even for 10-km distances and multi-user scenarios. We anticipate that our work will establish free-space networks as a viable solution for metropolitan applications and an indispensable complementary building block in the future global quantum internet.

## Introduction

The core functionality of a quantum communication network is to distribute quantum information between two or more parties[1]. Many revolutionizing applications of quantum networks have already been identified and a roadmap towards full-blown quantum internet has been proposed[2]. While there are still many uncertainties as to the technological platforms that will ultimately make up the quantum internet, one thing is clear: it will be a heterogeneous network of various special purpose sub-networks that employ different types of links and interconnects.

Quantum key distribution (QKD) networks have so far been the driving force for this development[3]. Although certainly not the only networks of interest, they have been paving the way also for other distributed quantum information processing protocols. For this reason, achievable secure key rates have often been used to benchmark the level of technological maturity of quantum networks in general. Currently, connecting many users distributed over large distances requires trusted nodes[4], which comes with the price of losing any possibility of end-to-end security. Unlocking the full quantum advantage in global-scale networks is therefore an ongoing challenge. A substantial proportion of current research is focused on extending the reach of individual quantum links, either through satellites[5–8] or fibre-based quantum repeaters[9–11].

Over shorter, metropolitan-scale distances, where end-to-end quantum state transmission is more easily achieved, the research focus is on a different set of challenges. One of the primary concerns is the question of scalability, i.e. how to increase the number of users in a network[12–15]. Another line of research aims to make metropolitan quantum network technology more accessible, flexible, and deployable[16–19]. An important challenge is also to interconnect networks based on different physical platforms[20]. However, all this progress is made almost exclusively with optical fibre links.

In some metropolitan application scenarios, end-to-end fibre links are not feasible. A possible alternative are terrestrial free-space links. These are still far behind the technological maturity of fibre-based systems, particularly in terms of the availability of off-the-shelf and plug-and-play solutions that are required for their deployment. Free-space links face additional challenges, such as link alignment, atmospheric turbulence, and daylight noise[21].

---

[*] Andrej.Krzic@iof.fraunhofer.de

[†] Fabian.Steinlechner@iof.fraunhofer.de

In the following, we argue that entanglement distribution is particularly well-suited to metropolitan-scale networks. Among the plethora of prepare-and-measure implementations, however, only a handful of groups has performed entanglement-based QKD over terrestrial free-space links[22–32], of which merely two were performed in daylight (see Supplementary Information for a detailed review). Note, that many of these experiments explicitly aimed at advancing technology towards satellite deployment, while the specific challenges of readily deployable terrestrial networks was rarely addressed. Secure key rates in the order of hundreds of bits per second (bps) have been achieved over kilometre distances at night[23,27], while in daylight, similar rates have been demonstrated over a distance of only 350 m[26]. It has therefore been widely argued that the entanglement-based approach lacks the technological maturity and applicability of alternative approaches[33–37].

Here, we make the case for a metropolitan free-space network architecture that can be deployed to secure communication at summits, conferences, and other events, or complement an existing network infrastructure whenever end-to-end fibre connections are not available. The architecture is built around a central entanglement server that streams entangled photons to users on the network. We developed the key building blocks of this architecture, including: a portable high-visibility entangled photon pair source, deployable and efficient free-space terminals that are specifically designed for metropolitan applications, and compact and passive quantum state analysis and detection with dedicated filtering for daytime operation. To demonstrate the competitiveness of this approach in terms of efficiency, we benchmark its key generation capacity in several link configurations. We performed QKD experiments in realistic metropolitan scenarios: over a 1.7-km link to a temporary container atop another building and between two government offices separated by 300 m, showcasing its capability of providing ad hoc quantum security. We demonstrated record kbps rates are achievable in nighttime as well as in broad daylight. Finally, we provide estimates of key rates for longer 10-km links and for scenarios with multiple free-space links.

### Entanglement-based quantum network

We consider a network architecture, where an entangled photon source acts as a server that streams entanglement into a metropolitan-scale network consisting of free-space links (Fig. 1a). The natural choice for placing the entanglement server (ES) is a central high-rise building with a clear line of sight to the relevant urban areas. Each of the end users owns a quantum receiver subsystem (QRS), which typically incorporates quantum state analysis and detection, timing and synchronisation electronics, and post-processing software. Although the network has a star topology at the physical layer (Fig. 1b), recently proposed multiplexing strategies can leverage the nonlocal correlations of entangled pairs to provide a fully connected mesh network at the quantum communication layer[12,15,19].

Employing quantum entanglement as the main network resource has several key advantages over the more established prepare-and-measure schemes. Entanglement offers additional layers of security via the prospect of device-independent protocols[38]. The ES can thus act as an untrusted relay node and lift the requirement for direct line of sight between the transceiver telescopes of any two users. Furthermore, the ES is all passive, meaning that the post-processing is delegated downstream to the user, which allows for standardization of ES, while the users can use an application-specific QRS. This gives a large degree of flexibility and even upgradeability, as more advanced schemes, protocols, and post-processing methods become available over time. Moreover, entanglement quite naturally supports many other promising applications beyond QKD[21].

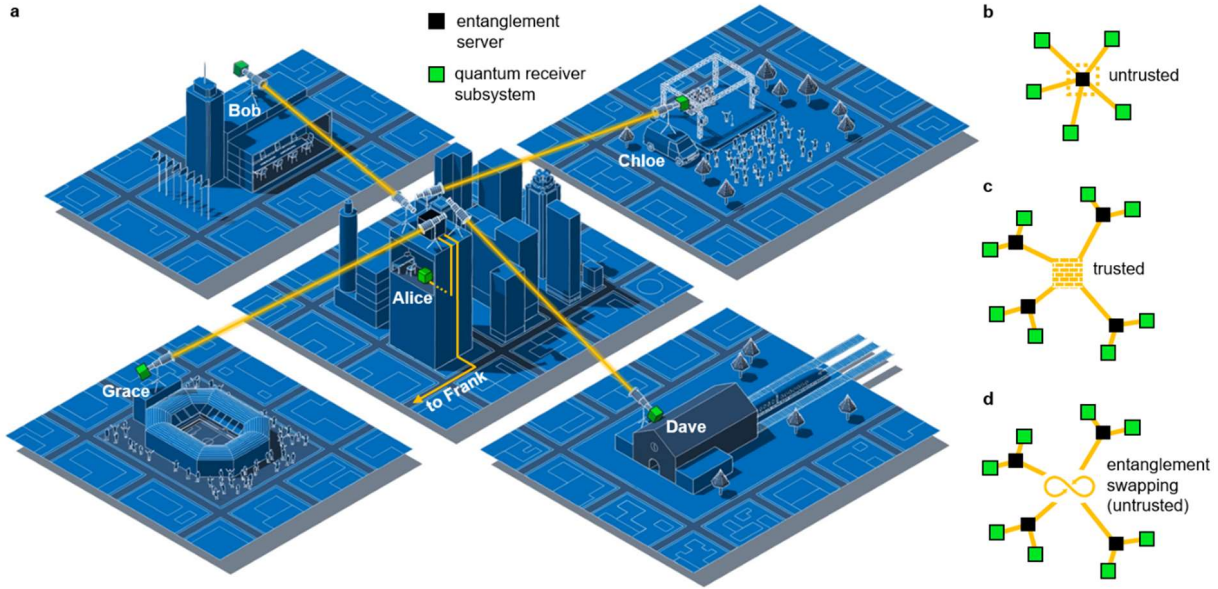

Fig. 1: Metropolitan entanglement-based free-space network. **a**, A standardized centrally located entanglement server (ES, black box) is streaming entangled photons into the network. Free-space channels are used to connect distant buildings and parts of a metropolitan area, while fibre connections may still be used in a complementary way, for example, to connect to offices within the central building. Each end user owns an application-specific quantum receiver subsystem (green boxes). **b**, The corresponding physical layer network topology. At the quantum communication layer, the network is a pairwise connected mesh, so that every end user can communicate with any other (not shown). **c**, A near-term extension possibility using several ESs and a central trusted node. **d**, Eventually, by introducing entanglement swapping, the trusted node could again be turned into an untrusted one, while the overall network topology would remain unchanged.

The network size and reach can be readily extended by introducing several ESs, all interconnected through a central trusted node and each distributing entanglement to end users in their own local areas (Fig. 1c). In a more distant future, the requirement for trusting the central node could eventually be lifted by introducing entanglement swapping[39], without changing the physical network topology (Fig. 1d). As indicated in Fig. 1a, the architecture is also not limited to free-space links – the ES can act as a convenient interface between a free-space and a fibre link segment, for example, to connect a metropolitan free-space network to the fibre backbone of a larger intercity network.

## Experimental setup

We developed a quantum communication system specifically suited for metropolitan applications. It consists of all the key building blocks necessary to implement the present free-space quantum network and can be fully deployed at a new site in less than a day, without requiring prior infrastructure apart from access to electricity. The system is schematically shown in Fig. 2 and resembles the Alice-Bob segment of the network in Fig. 1a. Our ES generates pairs of polarization-entangled photons at 810 nm (Methods), of which one is sent to Alice via a single-mode fibre, while the other one is sent to Bob using a free-space link. For fine alignment of the free-space link and beam stabilization, we use a 1064 nm beacon laser beam, which co-propagates along the signal photons from Alice to Bob. Analysis of the beacon beam at Bob provides a feedback signal for two fast steering mirrors, which in turn counteract fluctuations and stabilize the free-space channel in the presence of atmospheric turbulence and mechanical system instabilities. The use of a closed-loop beam stabilization system enables long-term operation and also facilitates the use of a spatial filter with a small field of view at Bob, which is necessary for daylight operation of the system (Methods).

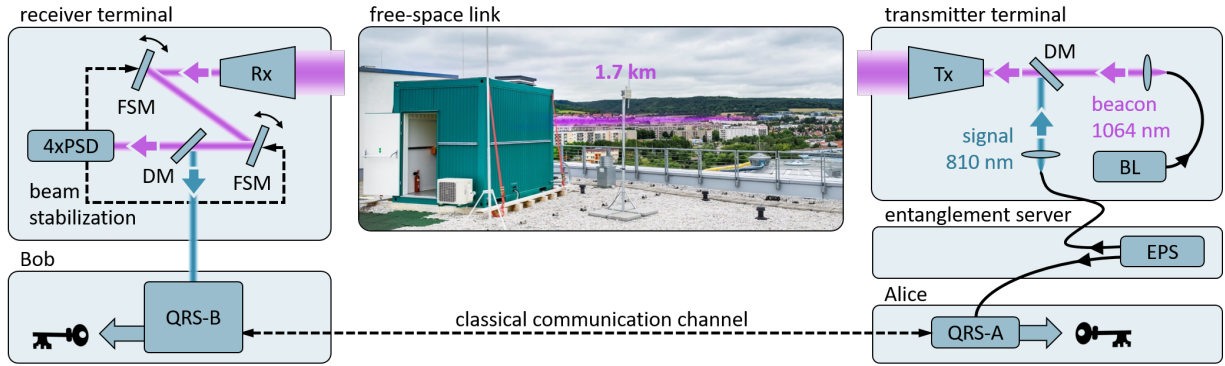

Fig. 2: Experimental setup. An entangled photon source (EPS) acts as a server that generates pairs of polarization-entangled photons at 810 nm. One photon of each pair is sent to Alice via a single-mode fibre, while the other one is sent to Bob via a free-space link spanned by a transmitter and a receiver telescope (Tx and Rx). A 1064 nm beacon laser (BL) creates a reference beam that is combined with the quantum signal at a dichroic mirror (DM) and co-propagates with it over the free-space link. At Bob, the beacon beam is analysed in a dedicated module consisting of four position sensitive detectors (4xPSD), which generates feedback signal for two fast-steering mirrors (FSM) that in turn stabilize the beam. Both Alice and Bob own a quantum receiver subsystem (QRS-A and QRS-B) for generating the secret key. Optical elements that are irrelevant for the present experiment have been left out for clarity here. The free-space link shown here is the 1.7-km link in Jena, Germany, between Fraunhofer IOF and Stadtwerke Jena.

Both Alice and Bob own a quantum receiver subsystem (QRS), which we developed for performing QKD with the BBM92 protocol[40] (Methods). Bob's QRS further incorporates specially designed spectral and spatial filtering modules for daylight operation. Alice's QRS, on the other hand, does not require dedicated filtering of daylight noise, since it is not directly exposed to the free-space link. Bob's QRS is also equipped with a motorized polarization controller that allows for automated alignment of the polarization frame of reference with Alice. The classical channel between Alice and Bob's QRS is realized with commercial radio antennas.

## Results

Using our system, we established a quantum link in Jena, Germany, between Fraunhofer IOF and a temporary container on top of a 1.7 km distant public service building. In Fig. 3, we benchmark the system performance in terms of quantum bit error rate (QBER) and achievable secure key rate (SKR) for two experiments that cover two extremes of link conditions: nighttime, when the background noise is negligible and the atmospheric turbulence weaker, and daytime, when solar radiation introduces considerable background noise and the turbulence is typically stronger[41]. For reference, we also show Bob's detected count rates, which are further split into signal and background noise contributions (Methods), as well as solar radiation measured by a weather station.

During the night experiment on 2 March 2022, we demonstrated a stable performance with QBER of less than 2%, resulting in an average SKR of 5.4 kbps over several hours. During the daytime experiment on 25 February 2022, however, the system performance varied to a much greater extent. At the beginning of this experiment, right before noon, the link was exposed to direct sunlight, i.e. no clouds were blocking the Sun as seen from the link (Supplementary Information). During these conditions that lasted roughly half an hour, the detected noise rates reached values larger than 400 kcps and were up to 3.8-times higher than the detected signal rates, despite strong spectral and spatial filtering. Nevertheless, the resulting QBER was kept below 3.4% and the SKR above 2.5 kbps, thanks to the further strong temporal filtering allowed by the tight temporal correlations of entangled photons. Over the next few hours, small clouds sporadically blocked the Sun, in turn reducing the background noise and restoring the system performance. Furthermore, we can see a clear correlation between the background photon rates and the independently measured solar radiation, and that the system performance varies as a direct consequence of sunlight. The only exception is the last drop of performance that happened after 14:00, which was not due to daylight but due to a sudden drop of the ES output, which is indicated by the simultaneous sharp drop of detected signal rates at Bob (Fig. 3) and at Alice (Supplementary Information).

We note that unlike for spatial filtering, the full potentials of spectral and temporal filtering were not exploited in daytime experiments. Optimizing the spectral filter bandwidth and coincidence window could substantially increase system resilience to daylight noise and increase the achievable secure key rates. The reported secure key rates are $\varepsilon$-secure with $\varepsilon < 10^{-10}$ (Methods), however, even security of $\varepsilon < 10^{-20}$ could be achieved by sacrificing only about 0.1 kbps of SKR (Supplementary Information).

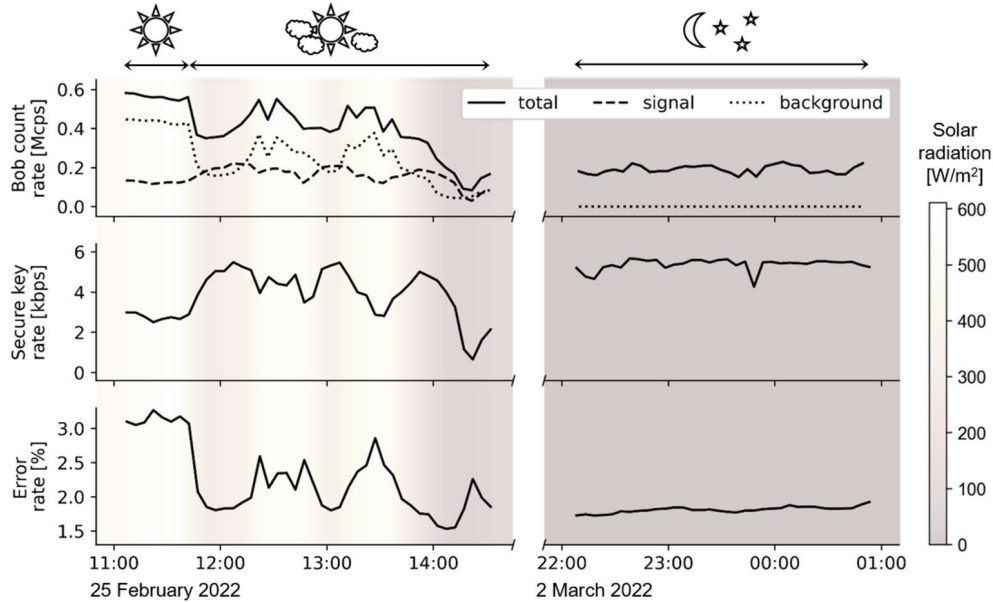

Fig. 3: Demonstration of free-space quantum key distribution. Single photon count rates at Bob, quantum bit error rates, and the corresponding secure key rates are shown for different link conditions. The daytime experiment (left) started around noon, with the link fully exposed to direct sunlight. It then continued with a sporadic decrease and increase of sunlight due to the moving clouds. During the night experiment (right), the background noise was negligible, resulting in a much more stable performance. The plotted values are 5-minute averages and the horizontal axis shows the local time of day. The background colour corresponds to the solar radiation measured independently by a weather station (solar radiation data: University of Applied Sciences Jena).

We extrapolated our measurements to estimate achievable secure key rates for longer links and multi-user scenarios (Methods). By taking the experimental data from our night experiment and considering additional loss in Bob's channel as estimated for a 10-km link, we arrive at SKR of 3.3 kbps. For a similar extrapolation of the daytime experiment data, scaling of the background noise rates on the distance would need to be further considered, which is beyond the scope of this work.

To gauge the performance in other link configurations, we simulate a scenario, consisting of two independent 1.7-km free-space links from the ES to Alice and Bob (Methods). Extrapolation of experimental data obtained during the night experiment shows that a SKR of more than 1.1 kbps would be possible. However, for the case of direct exposure of both links to sunlight, optimization of filtering parameters in all degrees of freedom would be required. For example, a 3-fold decrease of the spectral filter bandwidth to 1 nm would result in a SKR of more than 200 bps and optimization of the coincidence window offers further space for improvement. These estimates apply also to a general multi-user scenario based on wavelength division multiplexing, where each of the users is separated by an identical 1.7-km free-space link from the ES, assuming a perfect and uniform partition of the ES emission spectrum into the multiplexed channels (Methods).

Finally, we note that in July 2021, we deployed the system to establish a free-space QKD link between two 300 m separated government offices in Bonn, Germany, where we achieved similar key rates in night and in full daylight.

**Discussion and conclusion**

We proposed an entanglement-based free-space quantum network architecture as a viable solution for metropolitan scales. With a fully deployable QKD system, which is particularly suitable for ad hoc implementation in scenarios where fibre connections are infeasible, we demonstrated all the main building blocks of such a network. We achieved up to 5.7 kbps secure key rates over two different links and under various link conditions, with over 2.5 kbps in full daylight, indicating that the system has a potential for 24/7 operation. To our knowledge, this is the first demonstration of kbps-rate free-space QKD over kilometre distances with an entanglement-based approach. Moreover, it is an improvement of entanglement-based QKD daylight capability by an order of magnitude in both secret key rate and link range[26]. The results are also competitive with prepare-and-measure systems. With our large-aperture free-space terminals, specifically designed for metropolitan applications, we estimate kbps key rates are possible even over 10-km links. A full coverage of a city is therefore within reach with the present technology.

By implementing the quantum channel at 810 nm, we break with ongoing trends of transferring the free-space segments to telecom C band wavelengths. Network architectures implemented in C band typically strive to minimize the number of detectors[4,42], which is due to the fact that single-photon detection at 1550 nm is a cost driver. This does not apply in our case, where transmitter and receiver systems are of similar cost and complexity. However, 1550 nm will almost certainly remain the wavelength of choice for long fibre segments due to a substantially lower loss. A non-degenerate entangled photon source[43], which emits photons at 1550 nm and 810 nm, could therefore act as an elegant interface between fibre and free-space segments of a heterogeneous network.

There are many possible improvements to the present system on the horizon. Extending the two-user scenario to a fully connected multi-user network is straightforward with the latest multiplexing techniques[12,15,19]. By combining such a multiplexing scheme together with our system, we estimate kbps nighttime and hundreds of bps daytime rates are possible for a multi-user scenario with multiple 1.7 km free-space links. Sources that are capable of providing a large spectral bandwidth for multiplexing have been recently demonstrated, for example by Lohrmann *et al.*[44] Recent developments towards ultra-bright GHz rate entangled pair sources[44,45] show that a single entanglement server could support many end users.

Adaptive optics (AO) could significantly improve daylight performance by reducing the effective focal spot size and thus giving the possibility for tighter spatial filtering[34,37]. Furthermore, with efficient AO-enabled single-mode fibre coupling, hybrid free-space-to-fibre quantum links become possible. This would allow for physical separation of the end user from the receiver free-space terminal, introducing considerably more flexibility. Entanglement swapping[39] would not only increase the extent of the present network without the need for trusted nodes, but it would also enable integration into larger heterogeneous networks that employ fundamentally different forms of entanglement[20].

Another way forward is to exploit high-dimensional entanglement, which promises higher information capacity per photon, better security, and enhanced robustness to noise[46,47]. Although the practicality of orbital angular momentum (OAM) modes remains questionable for long-distance free-space communication[48], the OAM-entanglement might still be well suited for metropolitan distances[49]. AO would further improve key rates by protecting spatial mode entanglement from the influence of atmospheric turbulence[50]. An entangled photon source can also generate hyperentanglement – simultaneous entanglement in different degrees of freedom – quite naturally[51], which could be used to make a single server compatible with a range of receiver types. A recent experiment demonstrates that extending entanglement into the time-bin degree of freedom could also significantly boost secure key rates[52].

Finally, the distribution of entanglement lies at the heart of many applications beyond QKD. Practical and efficient metropolitan entanglement distribution networks would therefore facilitate the development of entirely new applications at the intersection of distributed sensing and quantum information processing, such as quantum clock synchronisation[53], long baseline interferometry[54], and multi-partite quantum cryptography[55,56].

## METHODS

### Entanglement server

Our entanglement server (ES) consists of an entangled photon source, which generates bipartite polarization entanglement. It is based on type-2 spontaneous parametric down conversion (SPDC) in a periodically poled potassium titanyl phosphate (ppKTP) crystal, employing an intrinsically phase-stable Sagnac configuration[57]. Pumping the crystal with a 405 nm laser generates an entangled pair state at 810 nm, which is of the form

$$|\Psi\rangle \propto |H_s\rangle|V_i\rangle + \beta e^{i\varphi}|V_s\rangle|H_i\rangle \tag{1}$$

where H and V are horizontal and vertical polarization states, and subscripts s and i stand for signal and idler paths. Coefficients $\beta$ and $\varphi$ are adjusted by manipulating the polarization of the pump to achieve a maximally entangled Bell state that we use for QKD.

During our experiments, the intrinsic photon pair generation rate (before any loss is considered) of the ES was estimated to be approximately 12 million pairs per second, over a spectral bandwidth of 0.45 nm (FWHM). Visibilities of up to 99.5% and 97.4% were measured in the low pump power limit for horizontal-vertical and diagonal-antidiagonal basis, respectively. The source was built on an optical breadboard and housed in a wheeled 19-inch rack.

### Quantum receiver subsystem

We developed two complete quantum receiver subsystems (QRS), one for Alice and one for Bob, for performing QKD based on the BBM92 protocol[40]. Each QRS consists of a standard polarization analysis module (PAM), 4 single-photon avalanche diodes (SPADs), time-tagging electronics, a rubidium clock, and units for post-processing. In the PAMs, we realized the randomized basis selection with a 50:50 beam splitter. We note, that this splitting ratio is not optimal and that by using a biased basis randomization, the final secure key rate could be further improved, albeit only slightly[28]. We used commercial single-photon avalanche diodes (SPADs) with detection efficiency of > 60%, dark count rate of < 500 cps, and timing resolution of 350 ps. To ensure matching of polarization frames of reference at Alice and Bob, a QRS also incorporates a waveplate-based polarization controller. For the night experiment, polarization basis alignment was performed right before the experiment, while for the daytime experiment, it was performed the night before, since the absence of background noise allowed for better initial alignment. See Supplementary Information for more information about the QRS.

Alice's QRS is almost identical to the QRS of Bob – the only difference is in its PAM, which is a much more compact and simplified version of Bob's. This is because Alice detects photons that are coming from the server via a single-mode fibre and is therefore not exposed to the free-space link environment and related challenges. Bob's PAM, on the other hand, incorporates additional spatial and spectral filters for daylight operation, as discussed in the next section. The complete QRS of Alice was housed in a wheeled 19-inch rack, while Bob's PAM was mounted directly onto the receiver free-space terminal and the rest of his QRS was housed in a rack.

### Daylight noise filtering

Noise suppression turns out to be of critical importance in a free-space channel, particularly during daytime, when atmospheric scattering of sunlight can cause the receiver telescope to collect noise photons at rates that are many orders of magnitude higher than the signal photon rates. To reduce the noise, we employ filtering in all three degrees of freedom: spectral, spatial, and temporal. Spectral filtering is realized with a stack of commercial interference filters, resulting in a 3 nm wide pass-band (FWHM) around the signal wavelength of 810 nm. The spatial filter consists of a Keplerian telescope with a tiny aperture in its focal plane, which has the effect of reducing the overal receiver system field of view (FOV)[58,59]. For our daylight experiments, we thus reduced the system FOV to 31 μrad, which was just large enough to transmit the majority of the signal, while blocking considerable amounts of noise. For comparison, this FOV is 2 orders of magnitude smaller than the FOV of our receiver telescope, resulting in a noise reduction by 4 orders of magnitude (noise rate scales with the square of the linear FOV), assuming a uniform distribution of noise over the FOV. For the nighttime experiment, the limiting field stop was the core of the multi-mode fibres in PAM, which resulted in the system FOV of about 52 μrad. Spectral and spatial filter modules were directly integrated into Bob's quantum receiver subsystem. Temporal filtering, on the other hand, comes with the coincidence-based detection. For both experiments here, coincidence window of 1 ns was used. Note, that this was near-optimal for the night experiment, however, its optimization in daytime could improve the achievable secure key rates. Furthermore, comparing our signal and filter spectra shows that we could employ at least a 3-times narrower spectral filter, reducing the noise by a factor of 3 while not considerably affecting the signal throughput.

### Free-space terminals

We developed two portable and highly versatile optical free-space terminals[60]. For the experiments presented here, we used the 810 nm channel for the quantum signal and the 1064 nm channel for the beacon laser, and the beam stabilization subsystem at the receiver terminal. Both terminals incorporate identical transceiver telescopes, which are based on an afocal three-mirror anastigmat (TMA) design combined with an additional fourth mirror to reduce the size and improve the optical quality. The off-axis telescope layout renders the transceiver system with high signal throughput due to the absence of a central obscuration. The shape of the individual mirrors is designed to provide diffraction-limited performance over the full telescope aperture of 200 mm and the entire FOV of 3.5 mrad. Furthermore, such a large aperture prevents severe signal loss due to diffraction and atmospheric turbulence even over link distances of several kilometres. At 810 nm, for our transmitter beam waist of 40 mm and following the well-established approach to model beam propagation through Kolmogorov turbulence[61], we estimate near-zero loss due to beam spreading (diffraction combined with turbulence induced spreading) over a 1.7 km link under medium turbulence with refractive index structure parameter $C_n^2 = 10^{-15}$ m$^{-2/3}$. For a 10 km link and the same $C_n^2$, we estimate the average loss due to beam spreading to still be as low as 2.1 dB (see Fig. 4), making these terminals particularly suitable for low-loss free-space links over metropolitan distances. For more technical details about the terminals, we refer the reader to [60].

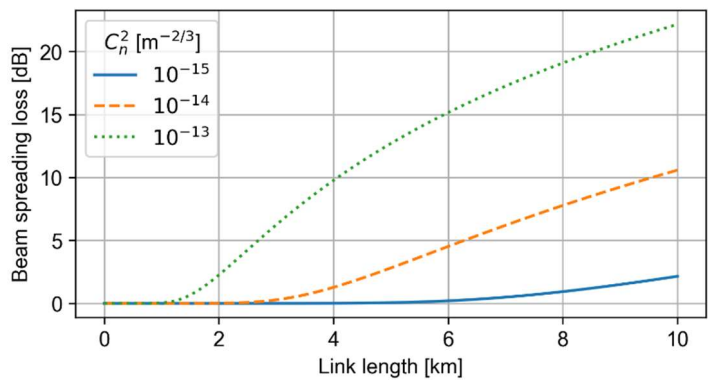

Fig. 4: Estimated link loss due to diffraction and turbulence induced beam spreading. The loss is shown for three different turbulence strengths, characterized by refractive index structure parameter $C_n^2$.

### Post-processing and secure key rate calculation

The first step of post-processing is to read data from the time-taggers. Alice and Bob exchange lists of time-tags and conduct time synchronization in two steps. First, a progressive cross-correlation-based pre-synchronisation algorithm with complexity $O(n \log n)$ is used to establish an initial clock offset over a wide range of multiple seconds. Once a coarse offset is found, fine synchronization works on a smaller block size that depends on the number of detections (Supplementary Information). For a more detailed discussion of similar approaches, we refer the reader to [62].

The sifting stage compares the measurement bases of the detections and keeps only events that share the same basis. The error estimation stage discloses a fraction of 10% of sifted key data in order to estimate the error rate in the quantum channel. Note, that the fraction of disclosed bits can be optimized to increase the amount of secure key, however, this was beyond the scope of experiments here. Information reconciliation (error correction) is then performed according to the cascade protocol[63] so that Alice and Bob agree on the same key with probability close to one. The subsequent confirmation stage tests whether information reconciliation was successful, i.e. it ensures the correctness of Alice's and Bob's key up to a minor error probability $\varepsilon_{\text{corr}}$ (the probability that Alice's and Bob's key differ and the confirmation stage does *not* detect it). For each key, Alice and Bob choose at random a hash function from a family of universal polynomial hash functions and calculate, exchange, and compare a hash of $t = 96$ bits of their reconciled key data. This results in $\varepsilon_{\text{corr}} = \frac{m-k}{t} 2^{-t}$, where $m$ is the number of bits after sifting and $k$ is the number of bits disclosed during error estimation.

For the experimental results reported here, all of the above was done online during the experiments. Secure key rates, taking into account also finite key effects, were then calculated offline using the security proof from [64]. The length of the secure keys is given as

$$l = \text{floor}\left[(m-k)\left(1 - h_2(\delta + \nu)\right) - r - t - 2a + 2\log_2(2\varepsilon_{\text{pa}})\right], \quad (2)$$

with

$$\nu = \sqrt{\frac{m(k+1)}{(m-k)k^2} \ln\left(\frac{2}{\varepsilon_{\text{pe}}}\right)}, \quad (3)$$

where $r$ is the number of bits disclosed during information reconciliation, $\delta$ is the error rate threshold for the parameter estimation test, $\varepsilon_{\text{pa}}$ is the accepted failure probability due to privacy amplification, and $\varepsilon_{\text{pe}}$ is the accepted probability that the actual error rate is larger than $\delta + \nu$, given that the parameter estimation test has been passed. We also introduced $a$, which corresponds to the length of the one-time pads used for encryption of authentication tags (the factor 2 represents one tag for each direction), since the original security proof assumes an authentic channel. We performed calculations on data blocks corresponding to 5-minute measurement times, with $\delta = 0.034$ and $a = 96$, and chose the total failure probability of the QKD protocol over an authentic channel $\varepsilon_{\text{qkd}} = 10^{-11}$. We further set $\varepsilon_{\text{pa}} = \varepsilon_{\text{qkd}} \times 10^{-3}$ and $\varepsilon_{\text{pe}} = \varepsilon_{\text{qkd}} - \varepsilon_{\text{pa}} - \varepsilon_{\text{corr}}$. The overall security of the calculated keys can be finally assessed by the parameter $\varepsilon$, which takes the authentication into account, and the fact that a part of the QKD keys is used for the authentication of the next round. $\varepsilon$ grows with the number of processed data blocks $n$ as[65]

$$\varepsilon \leq n(\varepsilon_{\text{qkd}} + \varepsilon_{\text{auth}}), \quad (4)$$

with $\varepsilon_{\text{auth}} = \frac{C}{a} 2^{-a}$, where $C$ is the length of the authenticated classical communication exchanged (in bits).

While these results were calculated offline, we note that the system is fully capable of live on-the-fly secure key generation. The system can implement the necessary steps, including delayed authentication according to [66] with polynomial universal hashing over GF($2^{96}$) with a pre-shared secret key, and privacy amplification using a Toeplitz matrix approach. The estimated rates reported here closely match the rate of secure keys generated live.

### System efficiencies

To calculate system efficiencies, we adapt the model from [67]. We first determine the coincidence-window-dependent detection efficiency

$$\eta_{\text{coin}} = \text{erf}\left[\sqrt{\ln(2)} \frac{\tau}{t_\Delta}\right], \quad (5)$$

where $\tau$ is the coincidence window and $t_\Delta$ is the full width at half maximum of the $g^{(2)}$ function, i.e. the correlation between Alice and Bob's detector clicks. Since $t_\Delta$ primarily depends on the time jitter of the detectors, we regard it as a constant of our system. We measured its value to be $t_\Delta = 710 \pm 8$ ps. Together with $\tau = 1$ ns, this results in $\eta_{\text{coin}}$ of about 83.8%. During the experiments, we were using 10% of the detection events for estimating Alice and Bob's single detection rate $S_{\text{det}}^a$ and $S_{\text{det}}^b$, respectively, and the coincidence detection rate $C_{\text{det}}$ (these events are excluded from the key calculation). We calculate the true coincidence rate as

$$C_{\text{true}} = \frac{C_{\text{det}} - C_{\text{acc}}}{\eta_{\text{coin}}}, \qquad (6)$$

where the accidental coincidence rate is estimated as

$$C_{\text{acc}} = S_{\text{det}}^a S_{\text{det}}^b \tau. \qquad (7)$$

Furthermore, by subtracting the dark count rate $S_{\text{DC}}^i$ and the background noise rate $S_{\text{BG}}^i$ from the detected single rates, we arrive at the contribution of the true (signal) photons,

$$S_{\text{sig}}^i = S_{\text{det}}^i - S_{\text{DC}}^i - S_{\text{BG}}^i, \qquad (8)$$

where superscript i ∈ {a, b} stands for Alice or Bob. We may then determine the total system efficiencies for Alice and Bob (sometimes called heralding efficiencies) as

$$\eta^a = \frac{C_{\text{true}}}{S_{\text{sig}}^b}, \qquad \eta^b = \frac{C_{\text{true}}}{S_{\text{sig}}^a}. \qquad (9)$$

These system efficiencies account for all the single channel losses from the moment a pair is created until it gets processed by the software, therefore it includes also detector efficiencies. The dark count rates are also regarded as system constants and we measured them to be $S_{\text{DC}}^a = (1{,}702 \pm 21)$ cps and $S_{\text{DC}}^b = (761 \pm 7)$ cps, on average.

For the night experiment, background count rate was negligible, therefore we can set $S_{\text{BG}}^a$ and $S_{\text{BG}}^b$ to zero. This allows us to use the model above to characterize $\eta^a$ and $\eta^b$. They are found to be very stable throughout the experiment, with the mean and standard deviation of $(8.90 \pm 0.03)$% for Alice, and $(1.63 \pm 0.08)$% for Bob. For the daytime experiment, we can estimate $\eta^b$ in the same way, since Alice's channel was fully fibre-based and thus experienced negligible background count rates. Doing so results in $\eta^b = (1.25 \pm 0.35)$%. On the other hand, Bob's background was far from negligible during this experiment, therefore we cannot extract $\eta^a$ in the same way. However, while $\eta^b$ depends on the link conditions that may considerably change throughout the day, $\eta^a$ is not expected to considerably change for days. We therefore characterized it the night before, arriving at $\eta^a = (8.13 \pm 0.09)$%.

The lower mean and the higher standard deviation of $\eta^b$ for the daytime experiment is not surprising, since narrowing down the spatial filter incurred additional loss and increased link efficiency fluctuations due to atmospheric turbulence. Note, that the source was realigned just before the night experiment, hence a slightly better $\eta^a$ then.

### Solar radiation and background count rate

To quantify sky brightness during our experiments, solar radiation data was extracted from the publically available database of the University of Applied Sciences Jena[‡]. The measurements were performed using a pyranometer, which measures total incident solar radiation over the whole hemisphere in the spectral band between 300 nm and 2.8 μm. The pyranometer was located approximately 1.2 km from Alice and approximately 2.6 km from Bob. Although not measured exactly where our experiments took place, and despite being measured over a much broader spectral range compared to our QKD spectral filter bandwidth, it still offers a reasonable metric for quantifying system-independent sky brightness at the link. This is evident from its corellation with the detected daylight noise (Fig. 3), which we estimate in the following way. As above, we determine the true coincidence rate $C_{\text{true}}$ from the detected rates. This time, however, $S_{\text{det}}^{\text{b}}$ also consists of a considerable contribution of $S_{\text{BG}}^{\text{b}}$ due to daylight. Using $\eta^a$ that was determined the night before, we can estimate $S_{\text{sig}}^{\text{b}}$ from $C_{\text{true}}$ and with it extract $S_{\text{BG}}^{\text{b}}$ from $S_{\text{det}}^{\text{b}}$, following the relations given above.

---

[‡] http://wetter.mb.eah-jena.de/station/index.html

## Performance estimation for a 10 km link

Extending the free-space link to longer distances would introduce an additional loss factor of η to Bob's signal channel. Using the secure key rate model from [67], it is straightforward to show that in the limit of much larger received signal rates compared to noise rates, key rates would scale linearly with η. During our night experiment, we achieved a mean secure key rate of 5.4 kbps, while the estimated mean total count rate at Alice and Bob was 1.0 Mcps and 190 kcps, respectively. During this experiment, total noise count was less than 2 kcps for Alice and less than 1 kcps for Bob, consisting mostly of dark count. For a 10-km link, where we estimate $\eta = 10^{-2.1/10} \approx 0.62$, as discussed above, we can therefore assume linear scaling of the key rate with η and thus expect 3.3 kbps key rate to be achievable with our system in nighttime.

## Extrapolation to multiple free-space link scenarios

To estimate achievable key rates in a dual free-space link scenario (two 1.7 km free-space links with ES in the middle), we take Bob's channel parameters and count rates extracted from our experiments and assume the same for Alice's channel. In particular, we first extract $\eta^b$ and $S_{\text{sig}}^{\text{b}}$ for our system, as explained above. Together with the directly measured $S_{\text{det}}^{\text{b}}$, we may then estimate true and accidental coincidence rate in such a scenario as

$$C_{true}^{(a\sim b)} = \eta^b S_{\text{sig}}^{\text{b}} \qquad (10)$$

and

$$C_{acc}^{(a\sim b)} = \left(S_{\text{det}}^{\text{b}}\right)^2 \tau. \qquad (11)$$

Imperfections in the ES, PAMs, and alignment of the polarization frames of reference at Alice and Bob lead to occasional erroneous polarization measurements[67]. We extract the rate at which this happens from the experimental data as

$$e_{\text{pol}} = \frac{C_{\text{err}} - C_{\text{acc}}/4}{C_{\text{sift}}}, \qquad (12)$$

where $C_{\text{err}}$ is the sum of measured coincidences in the wrong channels and $C_{\text{sift}}$ are the total measured coincidences after sifting. Finally, we assume a constant error correction efficiency of 1.1 (which is very close to the values achieved during our experiments) and estimate the key in the asymptotic limit according to the model in [67].

We can also extend these results to a multiple free-space link scenario based on wavelength division multiplexing, which was already demonstrated to work well in optical fibres[12,15]. Here, each of the $u$ users in the network establishes a secure key pairwise with every other user, resulting in $k = u(u-1)/2$ different secret keys generated in parallel. Since each pair of channels connecting any two users is independent from every other user pair, this means that a fraction $G/k$ of the intrinsic ES pair generation rate $G$ is used for generation of each key, assuming the ES emission spectrum can be uniformly distributed over the multiplexed wavelength channels for simplicity. Employing such wavelength division multiplexing and a $k$-fold increase of the ES pump power, we would therefore arrive at the same key rate predictions for a multiple free-space link case as in the dual free-space link case, assuming all the free-space links are identical.

## Acknowledgements

We thank Robert Jende, Ralf Steinkopf, Mathias Rohde, Sandra Müller, and Stefan Schwinde for their work on the transceiver telescopes; Herbert Gross for help with the spatial filter design; Mirko Liedtke and Carl Zeiss Microscopy GmbH for providing a motorized aperture for the spatial filter; Nico Döll for assembling the spatial filter; Emma Brambila-Tamayo and Rana Sebak for help with the entangled photon source; Stadtwerke Jena for giving us access to their rooftop for our experiments; Daniel Heinig for general experimental support; Rodrigo Gomez and Nina Leonhard for preliminary calculations; Aoife Brady and Claudia Reinlein for helpful discussions and support in the early stage planning; Markus Selmke and Julian Gritsch for administrative support; and Hanna Läkk for the network scheme figure graphics design. This research was conducted within the scope of the project QuNET, funded by the German Federal Ministry of Education and Research (BMBF) in the context of the federal government's research framework in IT-security "Digital. Secure. Sovereign." A.K. and C.S. are part of the Max Planck School of Photonics supported by the BMBF, the Max Planck Society, and the Fraunhofer Society. A.K. is co-sponsored by the European Space Agency (ESA) through the Networking Partnering Initiative (NPI) Contract No. 4000125842/18/NL/MH/mg (Project DIFFRACT).

# SUPPLEMENTARY INFORMATION

## Literature review

| Reference | FS link distance | Secure key rate | Duration | Day / night | Highlights |
|---|---|---|---|---|---|
| Peng et al. 2005[1] | 7.7 km + 5.3 km | 10 bps (mean) | 4 min | night | first demonstration of QKD over a dual FS link |
| Marcikic et al. 2006[2] | 1.5 km | 630 bps (mean) | ~10 h | night | first demonstration of QKD in real time |
| Ursin et al. 2007[3] | 144 km | 2.4 bps (mean) | 75 s | night | demonstration of suitability for satellite implementations |
| Erven et al. 2008[4] | 435 m + 1325 m | 85 bps (mean) | 6.5 h | night | first real-time QKD over two FS links |
| Peloso et al. 2009[5] | 350 m | night: 533 bps (max) day: 250-400 bps | 4 days | day & night | first daylight QKD |
| Erven et al. 2012[6] | 1.3 km | ~500 bps (mean) | 3 min | night | key rate improvement with a signal-to-noise ratio filter |
| Cao et al. 2013[7] | 7.8 km + 7.8 km | 0.4 bps (mean) | ~3 h | night | key rate improvement with biased basis choice |
| Yin et al. 2017[8] | low Earth orbit | 3.5 bps (mean) | 6 x 40 s (6 passes) | night | first satellite implementation |
| Yin et al. 2020[9] | low Earth orbit (dual downlink) | 0.12 bps (mean) | ~52 min (many passes) | night | first satellite implementation with a dual FS link |
| Ecker et al. 2021[10] | 143 km | 71.8 bps 300 bps (mean) | 68 s 15 s | night | system parameter optimization strategies for key rate improvement |
| Basso Basset et al. 2021[11] | 270 m | 2 bps (mean) | 224 min | night | first QKD with quantum dots |
| Mishra et al. 2022[12] | 200 m | 1.71 kbps (mean) | n.a. | night | first-time demonstration of kbps rates |
| Basso Basset et al. 2023[13] | 270 m | 11.5 bps (mean) | 3.5 days | day | first daylight QKD with quantum dots |
| Kržič et al. 2023 (this work) | 1.7 km | night: 5.7 kbps (max) day: > 2.5 kbps | 3-4 h | day & night | first-time demonstration of kbps rates over km-scale distances in night and day, and demonstration of deployable free-space QKD system for ad hoc metropolitan applications |

Table 1: Literature review. Reported free-space (FS) quantum key distribution (QKD) experiments with entangled photons are shown, together with the relevant aspects and highlights.

Here, we provide a review of reported successful quantum key distribution (QKD) experiments with entangled photons that were performed using free-space (FS) links (Table 1). All were performed between two users, employing signal wavelengths around 800 nm. While some demonstrate a dual FS link scenario[1,4,7,9], most were done over a single FS link. Achieved secure key rates heavily depend on the FS link distance. Over a 143 km terrestrial link, some 100 bps mean key rate was extracted for about a minute[10]. Note, that experiments over the particular 143-144 km link on Canary Islands[3,10] are quite obviously aimed at space applications. In fact, a large portion of all the experiments explicitly state satellite implementations as the aim and two actual satellite experiments have already been reported[8,9]. Experiments with more direct implications to metropolitan scenarios are a stable 10-h QKD with an average rate of 630 bps over a 1.5 km link in nighttime[2] and a daylight QKD with 250-400 bps over a 350 m link[5]. Quite recently, the first successful use of entangled quantum dots for FS QKD has been reported[11,13], however, the demonstrated secure key rates are still 1-2 orders of magnitude lower than for those that are based on the process of spontaneous parametric down-conversion (SPDC).

## Quantum receiver subsystem – details

Bob's quantum receiver subsystem (QRS) is schematically shown in Sup. Fig. 1. Signal photons that come from the server via a free-space link are analysed in a polarization analysis module (PAM). Here, polarization is projected into one of two mutually unbiased bases, chosen at random for each photon, which is realized with a 50:50 beam-splitter (BS). Photons that are transmitted by the BS are projected into the horizontal-vertical basis by the means of a polarization-beam-splitter (PBS) – we used high-extinction-ratio Wollaston prisms for that purpose. In the reflection arm of the BS, the additional half-wave plate rotates polarization of the photons by 45°, which in turn makes the subsequent PBS project them into the diagonal-antidiagonal basis. All the four outputs are finally coupled into multi-mode fibres and guided to the detectors. Each detector click is given a precise arrival time by a rubidium clock driven time-tagging unit. Time-tags from both sides are synchronised and processed by a dedicated post-processing software in real-time.

Finally, we need to ensure matching of polarization frames of reference at Alice and Bob. While each optical component along the path can in principle introduce some polarization transformation, by far the largest is due to the optical fibres that guide the photons from the ES. For this reason, Bob's PAM incorporates a polarization controller that consists of a stack of half- and quarter-wave plates in motorized rotation mounts. By rotating the wave plates, we can realize a transformation that inverts the net transformation of the optical elements and fibres. To find the inverse transform in practice, we insert a polarizer into the entangled photon source to send a well-defined reference polarization to Bob's PAM and then use the polarization controller to maximize the signal in the corresponding detection channel.

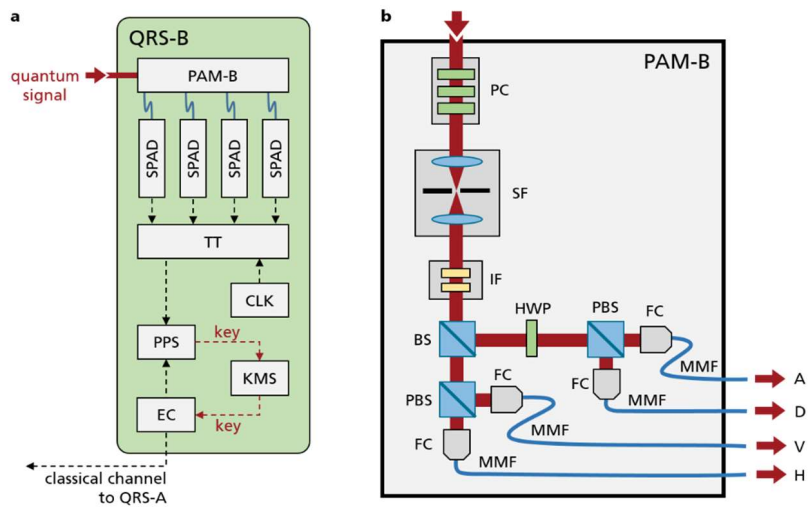

Sup. Fig. 1: Quantum receiver subsystem at Bob. **a**, The incoming free-space photons get sorted into 4 fibre channels according to their polarization state in the polarization analysis module (PAM-B) and detected with a single-photon avalanche detector (SPAD). Each SPAD click is given a precise arrival time by a rubidium clock (CLK) driven time-tagging (TT) unit. The post-processing software (PPS) processes the raw time-tags and generates encryption keys, which are fed to a key management system (KMS). From there, keys are pulled upon request for encryption (EC) of the classical communication channel between Alice and Bob. **b**, In PAM-B, a polarization controller (PC), consisting of a stack of waveplates, aligns Bob's polarization frame of reference with that of Alice. Excess noise is removed with a spatial filter (SF) and a stack of interference filters (IF). A 50:50 beam-splitter (BS) realizes a random choice of the measurement basis. Transmitted photons are measured in the horizontal-vertical (HV) basis by the means of a polarization-beam-splitter (PBS), while reflected photons are measured in the diagonal-antidiagonal (DA) basis by the means of a half-wave plate and a PBS. All the four outputs are finally coupled into multi-mode fibres (MMFs) using commercial fibre couplers (FCs).

## Fine synchronization

The fine synchronization uses a discrete histogram-based algorithm. Let $b$ be the bin size set to half the coincidence window and $N$ be the number of bins such that $Nb > o_{\max} - o_{\min}$, with $[o_{\min}, o_{\max})$ being the offset search range established in the pre-synchronisation or taken from the previous block. In the latter case, $o_{\min}$ and $o_{\max}$ are centred around the previous offset. Each bin $h[i]$ of the histogram accumulates the number of pairs, the sum and the sum of squares of their respective offsets in the range of $[o_{\min} + ib, o_{\min} + (i+1)b)$. Provided that the lower of the two (i.e. Alice and Bob's) data rates is similar or smaller than $1/Nb$, there will be at most $O(1)$ entries per time tag and thus the algorithm is linear in the number of time-tags. The distribution of coincidences is assumed to be the sum of a normally distributed signal and evenly distributed noise. The data in the histogram is used to estimate the parameters of this distribution. First, the noise level is estimated by using the bins away from the maximum, then the data around the maximum (±3 bins, i.e. 7 bins in total) is corrected for noise and used to estimate the signal parameters (peak offset, standard deviation, and number of correlated pairs). This allows to exclude obvious outliers in the estimate while achieving an accuracy way beyond the size of one histogram bin.

## Link exposure to sunlight

The link exposure to sunlight during our daytime experiments in Jena and in Bonn is shown in Sup. Fig. 2. For the daytime Jena experiment, we can clearly see that at about 11 am, when we started the measurements, the link was exposed to direct sunlight. By comparing the two pictures taken towards the Sun at different times, we can also see the clouds slowly moving towards the sun. About half an hour into the experiment, the group of clouds reached the Sun and started sporadically blocking it and thus reducing the sunlight at the link, as observed in the measured background count rates (Fig. 3 in the main text).

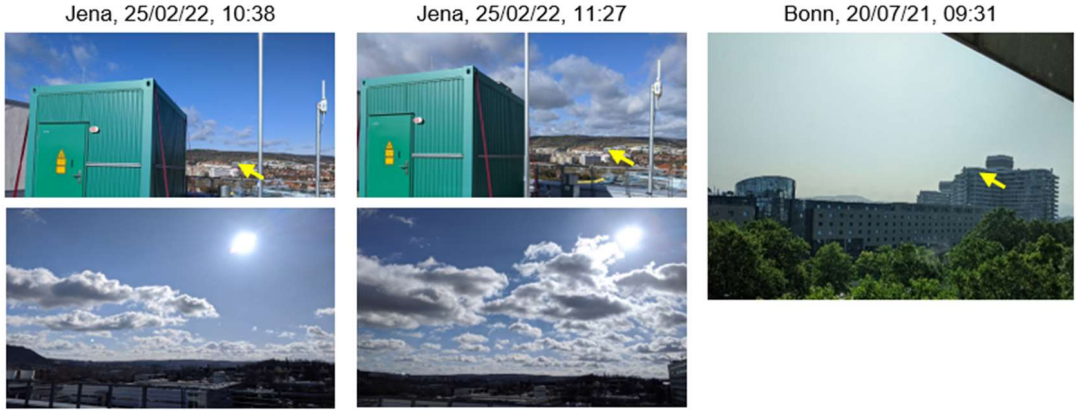

Sup. Fig. 2: Sunlight exposure during daytime experiments. Top three photos show the view from Bob towards Alice (the latter marked with an arrow) for the daytime experiments in Jena and Bonn. Bottom two photos show the link exposure to the Sun for the Jena daytime experiment, as seen from Bob.

## Alice count rate

The total single-photon detection count rate at Alice is shown in Sup. Fig. 3. The count rate was typically between 1.0 and 1.2 Mcps. Background and dark counts were in the order of a few kcps and are thus negligible. The sharp drop on 25 February 2022 after 14:00, which is also correlated with the drop of detected signal count rate at Bob (Fig. 3 in the main text), indicates that there was a considerable drop of the entangled pair source output. This lead to a significant drop of the system performance in terms of secure key rates that was unrelated to daylight.

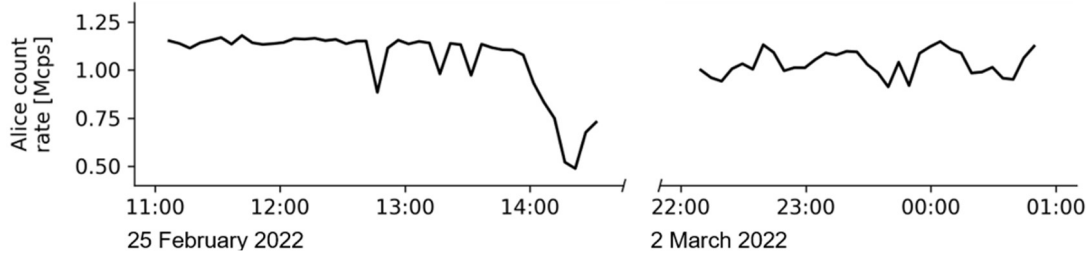

Sup. Fig. 3: Recorded single-photon count rate at Alice. The plotted values are 5-minute averages and the horizontal axis shows the local time of day.

## System performance dependence on security parameter

Table 2 shows mean secure key rates for different choice of security parameter $\varepsilon$ for two extremes: for the first part of the daytime experiment, when the link was exposed to direct sunlight, and for the nighttime experiment, when the background noise was negligible. We can see that even with security level of $\varepsilon < 10^{-20}$, kbps rates are achievable in nighttime as well as in direct sunlight.

|  | $\varepsilon < 10^{-10}$ | $\varepsilon < 10^{-15}$ | $\varepsilon < 10^{-20}$ |
|---|---|---|---|
| Direct sunlight | 2.78 kbps | 2.72 kbps | 2.67 kbps |
| Nighttime | 5.38 kbps | 5.30 kbps | 5.24 kbps |

Table. 2: Secure key rate dependence on security parameter $\varepsilon$. Mean secure key rates are shown for different $\varepsilon$ for two extreme cases of link conditions: direct sunlight and nighttime.